\documentclass[fleqn,usenatbib]{mnras}
\usepackage{newtxtext,newtxmath}
\usepackage{mathptmx}
\usepackage[T1]{fontenc}
\DeclareRobustCommand{\VAN}[3]{#2}
\let\VANthebibliography\thebibliography
\def\thebibliography{\DeclareRobustCommand{\VAN}[3]{##3}\VANthebibliography}
\usepackage{graphicx}
\usepackage{amsmath}
\usepackage{makecell}
\usepackage{array}
\usepackage{verbatim}
\usepackage{multicol}
\usepackage{longtable}
\usepackage{grffile}
\usepackage{textgreek} 
\usepackage{textcomp}
\usepackage{floatrow}
\usepackage{url}
\usepackage{array}
\newcolumntype{L}[1]{>{\raggedright\let\newline\\\arraybackslash\hspace{0pt}}p{#1}}
\newcolumntype{x}[1]{>{\centering\arraybackslash\hspace{0pt}}p{#1}}
\usepackage{siunitx}
\usepackage[parfill]{parskip} 
\graphicspath{{Figures/}}
\usepackage{verbatim} 
\usepackage{fancyvrb}
\usepackage{adjustbox}
\usepackage{makecell}
\usepackage{setspace}
\usepackage[font=small]{caption}
\usepackage{threeparttable}
\usepackage{csvsimple}
\usepackage{subcaption}
\usepackage{pdfpages}
\usepackage{verbatim}
\usepackage{afterpage}
\usepackage{scalerel}
\usepackage{tikz}
\usetikzlibrary{shapes.geometric, arrows}
\usepackage{dirtytalk}
\usepackage{multirow}
\usepackage{rotating}
\usepackage{adjustbox}
\usepackage{pdflscape}
\usepackage{graphicx}
\usepackage{float}
\usepackage{gensymb}
\usepackage{kantlipsum}
\usepackage{mathtools}
\usepackage{caption}
\usepackage{verbatim}
\usepackage{xr}
\usepackage{subcaption}
\usepackage{glossaries}
\usepackage{wrapfig}
\usepackage[font={normal}]{caption}
\usepackage{amsfonts} 
\usepackage{graphicx} 
\usepackage{xspace}
\usepackage{blindtext}
\def\Plus{\texttt{+}}

\graphicspath{{figures/}}
\usepackage{enumitem}
\usepackage{cprotect}
\DeclareUnicodeCharacter{2212}{-}
\title[Radio jets in NGC\,1068]{Radio jets in NGC\,1068 with $e$-MERLIN and VLA: structure and morphology}
\author[I. M. Mutie et al.]{Isaac M. Mutie,$^{1,2}$ \thanks{E-mail: mumoisaac@gmail.com}
David Williams-Baldwin,$^{2}$
Robert J. Beswick,$^{2}$
Emmanuel K. Bempong-Manful,$^{2,3}$
\newauthor{Paul O. Baki,$^{1}$
Tom W. B. Muxlow,$^{2}$
Jack F. Gallimore,$^{4}$
Susanne E. Aalto,$^{5}$
Bililign T. Dullo,$^{6}$}
\newauthor{and Ranieri D. Baldi$^{7}$}
\\
$^{1}$Department of Astronomy and Space Science, Technical University of Kenya, P.O Box 52428 $-$ 00200, Nairobi, Kenya\\
$^{2}$Jodrell Bank Centre for Astrophysics, School of Physics and Astronomy, The University of Manchester, Manchester M13 9PL, UK\\
$^{3}$School of Physics, University of Bristol, Tyndall Avenue, Bristol BS8 1TL, UK\\
$^{4}$Department of Physics and Astronomy, Bucknell University, Lewisburg, PA 17837, USA\\
$^{5}$Department of Space, Earth and Environment, Chalmers University of Technology, 412 96 Göteborg, Sweden\\
$^{6}$Departamento de Física de la Tierra y Astrofísica, IPARCOS, Universidad Complutense de Madrid, E-28040, Madrid, Spain\\
$^{7}$Instituto di Radioastronomia - INAF, Via P. Gobetti 101, I-40129 Bologna, Italy}
\date{Accepted XXX. Received YYY; in original form ZZZ}

\pubyear{2023}

\begin{document}
\label{firstpage}
\pagerange{\pageref{firstpage}--\pageref{lastpage}}
  \maketitle

\begin{abstract}
We present new high-sensitivity $e$-MERLIN and VLA radio images of the prototypical Seyfert 2 galaxy NGC 1068 at 5, 10 and 21\,GHz. We image the radio jet, from the compact components NE, C, S1 and S2 to the faint double-lobed jet structure of the NE and SW jet lobes. We map the jet between, 15\,k$\lambda$\,$-$\,3300\,k$\lambda$ spatial scales by combining $e$-MERLIN and VLA data for the first time. Components NE, C and S2 have steep spectra indicative of optically-thin non-thermal emission domination between 5 and 21 GHz. Component S1, which is where the AGN resides, has a flat radio spectrum. We report a new component, S2a, a part of the southern jet. We compare these new data with the MERLIN and VLA data observed in 1983, 1992 and 1995 and report a flux decrease by a factor of 2 in component C, suggesting variability of this jet component. With the high angular resolution $e$-MERLIN maps, we detect the bow shocks in the NE jet lobe that coincide with the molecular gas outflows observed with ALMA. The NE jet lobe has a jet power of $P_{jet-NElobe}\,=\,$6.7\,$\times$\,10$^{42}$\,erg\,s$^{-1}$ and is considered to be responsible for driving out the dense molecular gas observed with ALMA around the same region.
\\

\end{abstract}

\begin{keywords}
galaxies: active -- galaxies: individual: NGC 1068: jets -- cloud interactions: radio continuum: general -- radiation mechanisms: synchrotron. 
\end{keywords}



\section{Introduction}

Jets and outflows from active galactic nuclei (AGN) interact with the surrounding medium in their host galaxies. These interactions have far-reaching consequences on their immediate environments and affect co-evolution of supermassive black holes (SMBHs) and their host galaxies. On galactic scales, the interaction between AGN and interstellar medium (ISM) is known to release enormous amounts of energy into the surrounding environment in the form of radiation, collimated relativistic plasma (jets) and winds \citep{Planesas_1991,AGN_starb_Londsdale_1993,agn_winds_zubovas2012}. Consequently, AGN feedback has an impact on the small scale environment of the host galaxy by preventing ambient gas from cooling, quenching star formation \citep{feedback_ishibati_2012,feedback_fabian,feedback_raffaella}, or driving away gas by creating massive outflows \citep{Ishibashi-2016MNRAS,outflow_agn_harrison_2018,Torrey-AGNfeedback}.

High angular resolution radio observations provide unique insights because the radio waves travel extinction free, making them a useful probe for understanding source structure and morphology, as well as the nature of the interaction between the radio emitting jets and the surrounding medium. High fidelity radio imaging of the collimated jet emission provides important information on the jet power, orientation and interaction with the material in the host galaxy.

One of the best known nearby and radio-jetted AGN that has been studied across the electromagnetic spectrum is NGC 1068. It is classified as a Seyfert\,2 galaxy \citep{Khachikian&Weedman1974} and has been regarded as a prototype of this class \citep[e.g.,][]{koski1978}. The distance to NGC~1068 is 14.4\,Mpc at $z$ = 0.00379 \citep{Tully_1994,capetti_HST_1995,Bland-Haw_etal_1997}. This distance corresponds to 72 pc per arcsecond. The radio jet in NGC 1068 has been studied extensively previously (e.g. \cite{Muxlow1996,Gallimore_1996GTrue}), and can be divided into three regions: the central region (components NE, C, S1, S2 and S2a), the north-east (NE) region (NE plume, NE lobe) and the south-west (SW) region (SW plume, component S3 and SW lobe). In other wavebands, the AGN in NGC 1068 has also been studied in the X-rays \citep{chadra_xraY_young_2002,wang-x-ray-nucleus-2012ApJ} and neutrinos have been detected in the direction of NGC 1068, possibly from coronal activity within the AGN \citep{Inoue-neutrino-radio_2020,neutrino-icecube-2022}. Recent Atacama Large Millimeter/submillimeter Array (ALMA) dense molecular gas observations of NGC 1068 by \citet{Garcia-Burillo2014,Garcial_alma_imaging_2017A&,garcia-2019-core,gallimore-co-2016ApJ,imanishi-2016-gas,Gas_Imanishi_2018} and \citet{ Impellizzeri_2019}, have revealed the details of the jet$−$ISM interactions in this gas rich galaxy. Therefore, multi-wavelength studies of NGC 1068 are important in understanding the interplay of the AGN, jet and ISM properties in this source.

In this work, we image the jets in NGC 1068 between 60\,mas with the $enhanced$-Multi Element Radio Linked Interferometer Network ($e$-MERLIN) to $\sim$\,500\,mas resolution scales with the Karl G. Jansky Very Large Array (VLA). We investigate the structure and morphology of the inner radio jets in NGC 1068 from pc to kpc spatial scales. Our observations build on those from \cite{Muxlow1996} and \cite{Gallimore_1996GTrue} where data from the predecessor VLA and MERLIN interferometers were used. These two interferometers have since undergone significant upgrades to their receivers and correlation capabilities. The new wideband receivers provide improved $uv-$coverage and sensitivities. The overlapping $uv-$coverages between our $e$-MERLIN and VLA enable us to combine them, creating a synthesised interferometer, with angular resolution of $e$-MERLIN and VLA sensitivity. These combinations are done in this work for the first time of any NGC 1068 radio data. 

In this work, we aim to investigate the structure and morphology of all the components of the inner radio jets in NGC 1068. To this aim, we determine the radio jet power and study the effect of the AGN feedback on the surrounding ISM. We also investigate the impact that the ISM has on the jet structure and orientation on pc to kpc spatial scales. This detailed study of the radio jets in NGC 1068 is important to understand how energy from SMBHs is dissipated to the rest of the galaxy through radio jets. In Section  \ref{observations} we present observations and data reduction. In Section \ref{results} we present our results. In Section  \ref{discussion} we discuss the findings of this work.

\section{Observations and data reduction}\label{observations}
\subsection{The \texorpdfstring{$e$-MERLIN 5 GHz (C band) data}{Lg}} \label{data-emer-c-2018}
NGC 1068 was observed at a central frequency of 4.7\,GHz (C band) with a bandwidth of 512\,MHz across three epochs on 2018 January 7, 13 and 14 (MJD 58125, 58131, 58132) under project ID: CY6216. The observations are summarised in Table \ref{table_observations}. Each observation had four spectral windows (spws) across the 512 MHz bandwidth (128 MHz per spw). Each spw had 512 channels, but this was averaged to 32 channels as part of the processing procedures. Scans of the target source were interleaved with the phase calibrator, J0239-0234. 3C286 and OQ208 were also observed as the flux and bandpass calibrators, respectively. Target and phase calibrator scans were six and two minutes long, respectively. Each epoch consisted of a total of eight hours of cycling between the target and phase calibrator.

We inspected the data and after flagging any obviously bad data or radio frequency interference (RFI), we plotted data in Common Astronomy Software Applications (\verb .casa.) \verb. plotms. \citep{casa_McMulllin2007} and inspected it in time and frequency, per antenna, baseline and spectral window. Outlier points were flagged. We calibrated the data using the $e$-MERLIN \verb .casa. pipeline version: v1.1.08 (\citealt[eMCP]{eMCP-javier}) using  \verb .casa. version 5.5.0. We then performed flux, bandpass and phase calibration along with additional flagging of bad data, and preliminary imaging of the target and phase calibrator fields \citep{eMCP-javier}\footnote{\url{https://github.com/e-merlin/eMERLIN_CASA_pipeline}}. The calibrated pipeline data products were further averaged in time and frequency to 8 seconds and 4 MHz channels prior to self calibration.

\begin{figure*}
    \centering
    \includegraphics[width=18cm]{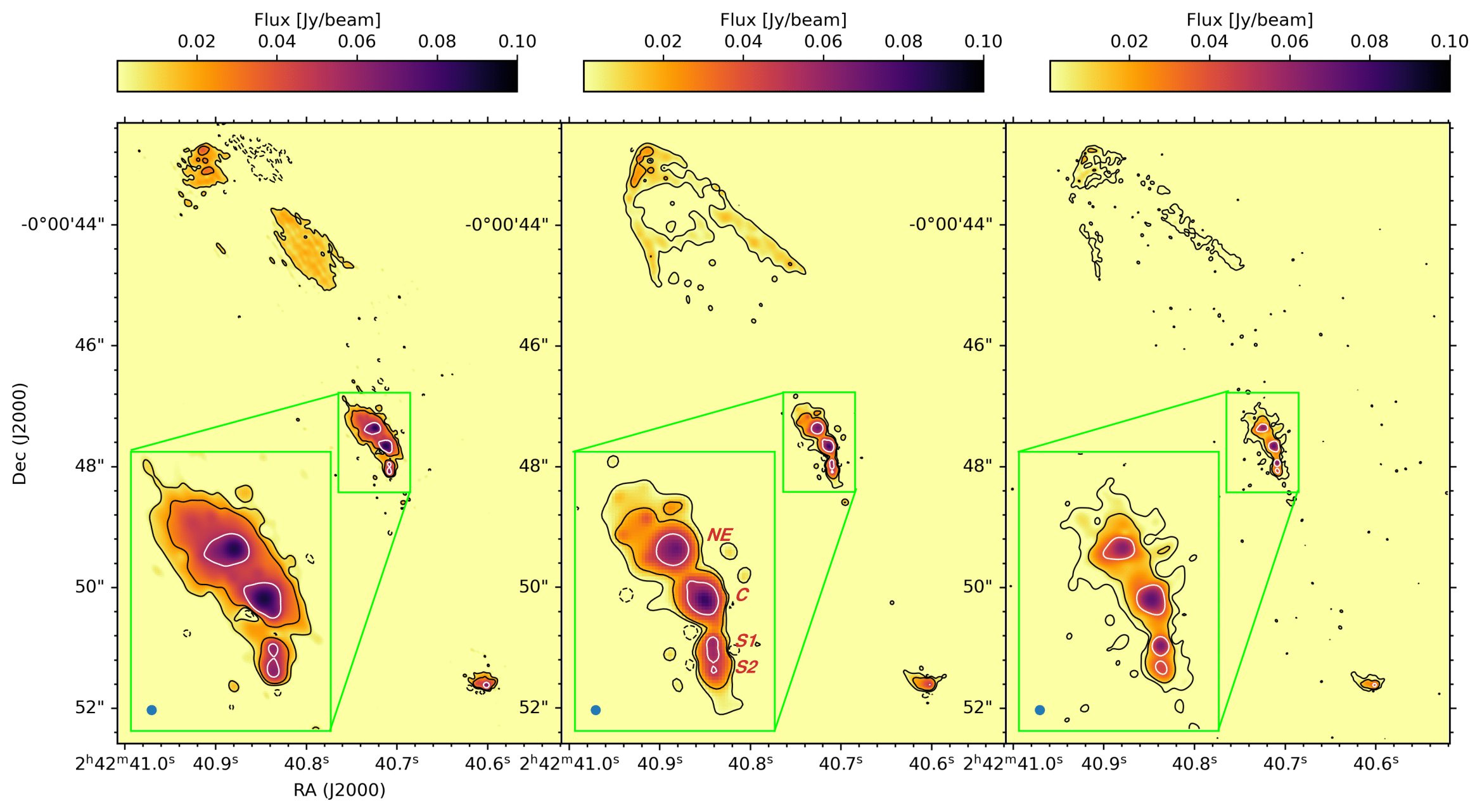}
    \caption{Left panel -- The $e$-MERLIN 5 GHz images of NGC 1068. Black contours show emission at [-3, 3 and 9]$\sigma$, white contours show 60$\sigma$ emission. $\sigma=95\micro$Jy and refers to the rms noise. Middle panel -- VLA 10 GHz image of NGC 1068. Black contours show emission at [-3, 3 and 9]$\sigma$, white contours show 75$\sigma$ emission, $\sigma=60\micro$Jy. Right panel -- VLA 21 GHz image of NGC 1068. Black contours show emission at [-3, 3 and 9]$\sigma$, white contours show 55$\sigma$ emission, $\sigma=42\micro$Jy. In all images, negative contours are dashed. Beams are represented by the blue ellipses in the bottom left corner of each image. The images have been created at matching $uv-$spacings of 65$-$2600 klambda. The three images are restored with a common circular beam of 0.05 arcsec.}
    \label{fig:c_k}
\end{figure*}

We imaged and self calibrated the data using the \verb .casa. task \verb .tclean. with cell size of 0.01 arcsec (about a fifth of the $e$-MERLIN angular resolution) to Nyquist sample the data. We used the multi-term multi-frequency synthesis deconvolver (mtmfs) deconvolver with Taylor polynomial ($nterms$) of 2 to maximise on modelling the wideband sky brightness distribution. To maximise the angular resolution, we used a robustness parameter of \,$-$2 in Briggs weighting \citep{briggs-weighting-1995}, resolving the compact components as shown in Fig. \ref{fig:c_k} (a).  

\subsection{The VLA 6, 10 and 21 GHz (C, X and K band) data} \label{vla-4-12-data}

We obtained four observations with the VLA\,-\,A array at 21 GHz (K band). These observations were made on 2015 June 26, 28, July 3, 11 (MJD 57199, 57201, 57206, 57214) under project code 15A-345. The VLA 6 and 10 GHz data were retrieved from the VLA Archive\footnote{\url{https://data.nrao.edu/portal/}}. These data were observed on 2021 January 20 and 26 (MJD 59234 and 59240) under project ID: 38677764. The two observations were performed at 8-12 GHz (X band) and 4-8 (C band) GHz, respectively. In all observations, the target was interleaved with the phase calibrator, J0239-0234, with 3C138 used as the flux and bandpass calibrator.

We calibrated all the VLA data using the standard VLA \verb .casa. pipeline, version: 6.4.1. The pipeline data reduction steps include observation file conversion (SDM-BDF to MS), flagging and flux, bandpass and phase calibrations. The National Radio Astronomy Observatory (NRAO) website\footnote{\url{https://science.nrao.edu/facilities/vla/data-processing/pipeline}} provides more detailed descriptions of the pipeline reduction steps. In each observation we split the target source from the calibrated dataset then averaged in time (8 seconds) and frequency (4 MHz). We then combined observations with multiple epochs using \verb .casa. task \verb .concat.. 
The target source was self calibrated with progressively smaller solution intervals, starting from the full scan length of 480 seconds to 8 seconds. The phase solutions converged to within 2 degrees. A final amplitude-phase self calibration was performed and solutions applied.

\begin{table}
\resizebox{\columnwidth}{!}{\begin{tabular}{ c c c c c c}
\hline
 Synthesis & Bandwidth&Central& Time on & Sensitivity & Angular  \\
 telescope & (GHz) & frequency & source & (rms) & resolution  \\
 &  & (GHz) &(hrs) & ($\micro$Jy\,beam$^{-1}$) & (arcsec) \\
 \hline\hline
 $e$-MERLIN & 4.4$-$4.9  & 4.7 & 18  & 60 & 0.05  \\
 \hline
 VLA$-$A$^{\text{C}}$ \newline &4$-$8  & 6  & 0.75 & 27 & 0.27 \\
 \hline
 VLA$-$A$^{\text{X}}$ \newline &8$-$12  & 10  & 0.75 & 65 & 0.16 \\
 \hline
 VLA$-$A$^{\text{K}}$ & 18.5$-$23.5  & 21 & 3  & 17 & 0.08 \\
 \hline
\end{tabular}}

\caption{Summary of the four observations from $e$-MERLIN and VLA. Superscripts C, X and K show VLA C$-$X bands at central frequencies as indicated. Where applicable, the measurements were taken by Gaussian fitting. The images used robust $-$2 with Briggs weighting in all cases.}   
\label{table_observations}
\end{table}

\begin{figure}[!ht]
\centering
    \subfloat[\centering]{{\includegraphics[width=8cm]{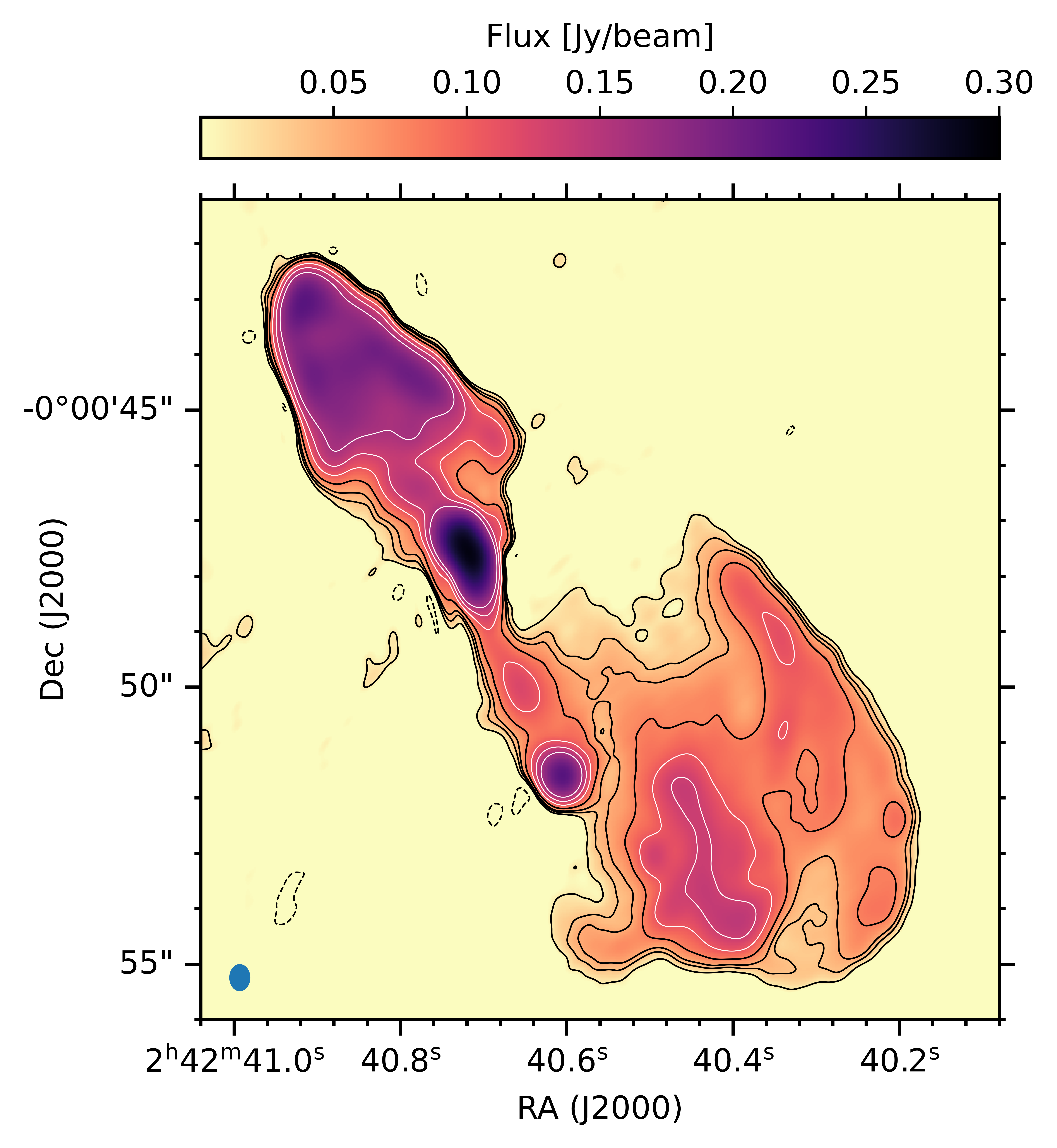}}}%
    \qquad
    \subfloat[\centering]{{\includegraphics[width=8cm]{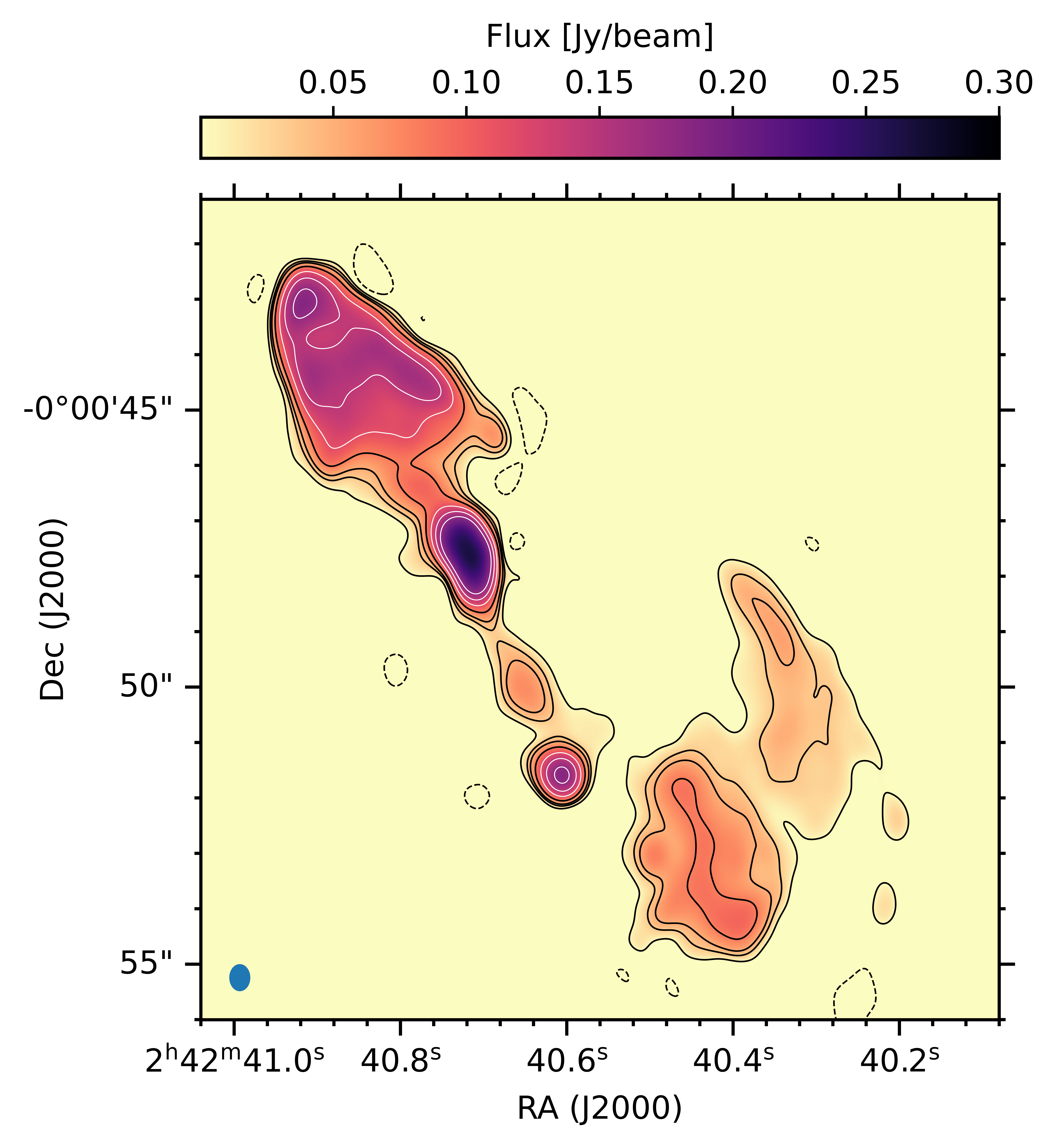}}}%
        \caption{(a), C band (6 GHz) image of VLA. The image is restored with $0.48\,\times\,0.39$ arcsec$^{2}$ beam (the blue patch in the bottom left corner), just like in \citet{Gallimore_1996GTrue}. Black contours show emission at [-4, 5, 10, 15, 30]$\sigma$ where sigma$=27\micro$Jy. White contours show emission at [64, 128 and 254]$\sigma$. Negative contours are dashed. (b), VLA X (10 GHz) band image with $\sigma=65\micro$Jy. All other parameters are the same to (a). }
    \label{fig:vla-C-X}
\end{figure}

\begin{figure}[!ht]
    \centering
    \includegraphics[width=8cm]{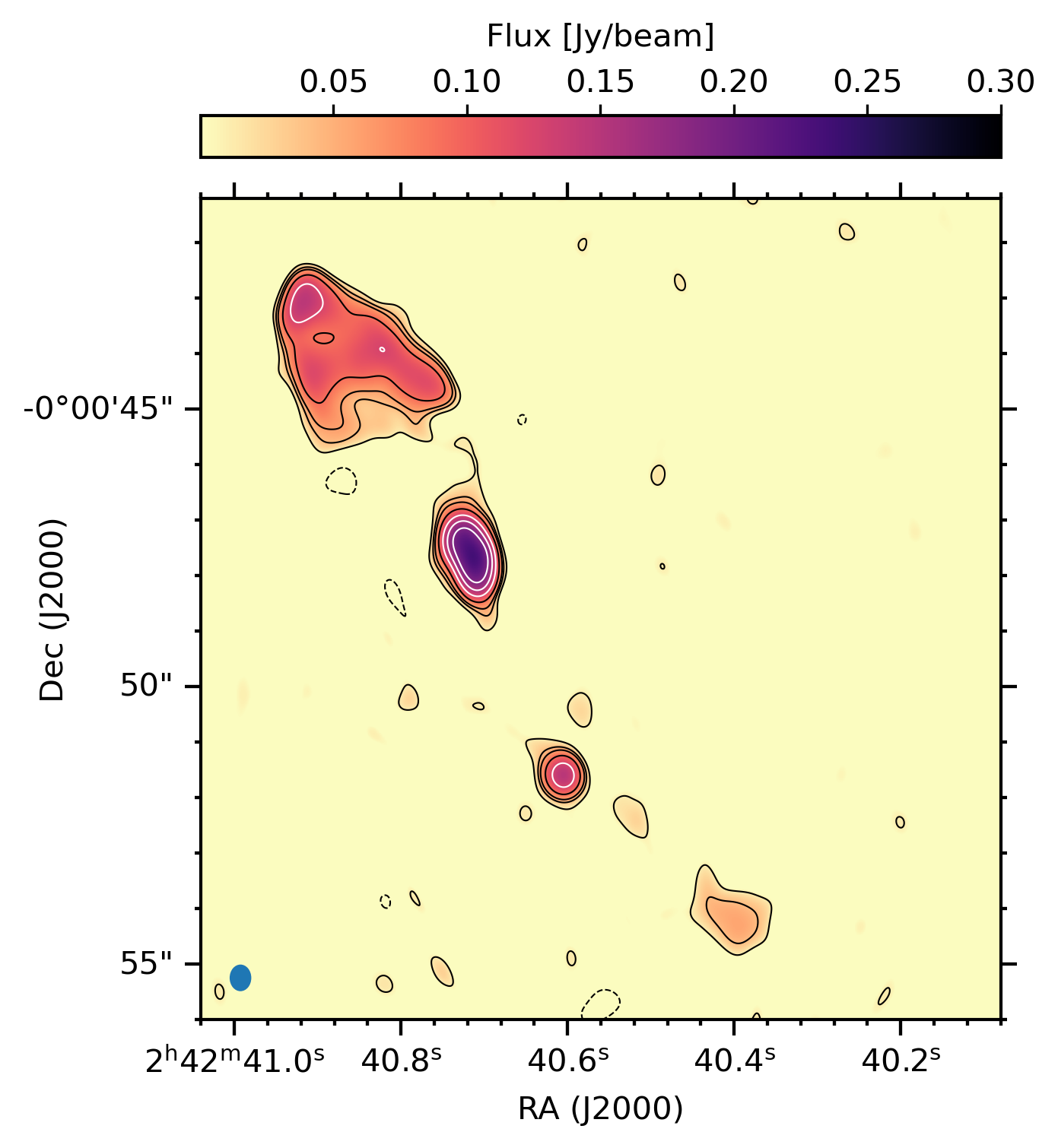}
    \caption{A combined VLA K band 2 (18.5$-$23.5 GHz) image of NGC 1068 at a central frequency of 21 GHz. The image is restored with a $0.48\,\times\,0.39$ arcsec$^{2}$ beam (the blue ellipse in the bottom left corner). Black contours show emission at [-4, 5, 10, 15, 30]$\sigma$ where sigma$=9.9\micro$Jy. White contours show [64, 128 and 254]$\sigma$ emission. Negative contours are dashed.} 
    \label{fig:vla-CX}
\end{figure}

We imaged the data using the \verb .casa. task \verb .tclean.. We used a cell size of 0.04 and 0.02 arcsec for the VLA 4$-$12 GHz and 21 GHz VLA-A array data, respectively. The mtmfs deconvolver was used with Taylor polynomial ($nterms$) of 3 to model the wideband sky brightness distribution, removing persistent artefacts and radiating spokes that occur with the VLA data near bright sources. To achieve higher angular resolution, a robustness parameter of $-$2 was used with Briggs weighting \citep{briggs-weighting-1995}. The images are shown in Fig. \ref{fig:c_k} (b) and (c) for VLA 10 and 21 GHz respectively. A larger restoring beam of 0.49\,$\times$\,0.38 arcsec$^{2}$ was used in order to map the jet's faint structure (see Figs. \ref{fig:vla-C-X} and \ref{fig:vla-CX}). 

In the course of this work, we serendipitously discovered radio supernova SN2018ivc about 18 arcsec on the north-east of the AGN \citep{mutie-sn2018ivc}. The supernova resides in the northern inner spiral arm of NGC 1068 and along the SB bar \citep{thronson-sb-bar-1989ApJ,Gallimore_1996GTrue}. This was the first radio detection of the source. The point source was missing in our VLA 21 GHz 2015 data, but appeared in VLA 2021 6 and 10 GHz data. Our April/May 2022 observations of NGC 1068 with $e$-MERLIN confirmed the source was present, while it was missing in January 2018 $e$-MERLIN observations (see Section \ref{data-emer-c-2018}). We included SN2018ivc in our model in the most recent datasets to improve our overall self calibration. Follow-up observations have been made $e$-MERLIN and the results will be discussed in future works.


\subsection{Data Combination, \texorpdfstring{$e$-MERLIN$\Plus$VLA}{Lg}}

In order to image the full jet structure at all spatial scales from the core to the lobes, we combined the VLA and $e$-MERLIN data from 4 to 12 GHz. The VLA has a maximum baseline of 36 km, giving an angular resolution of 0.2 arcsec, corresponding to $\sim$\,14.4 pc. This helps to detect the large-scale emission in the jet lobes. $e$-MERLIN's long baselines of up to $\sim$\,217 km yields angular resolution of $\sim$\,0.05 arcsec, corresponding to $\sim\,$4 pc, and resolves the compact components of the jet. 

The shortest $e$-MERLIN spacings overlap with the longest \mbox{VLA-A} array spacings, allowing data combination. Interferometric data combination has been successfully performed in the past for VLA and $e$-MERLIN (e.g. \citealt{beswick-3c293-2002, williams-ngc6217-2019,emmanuel-jets-2020,baldi-combined-2021}). Both the $e$-MERLIN and VLA data were re-weighted using the \verb .statwt. task in \verb .casa.. We combined the data using \verb .casa. task \verb .concat., where we down-weighted the VLA data by a factor of 3 compared to $e$-MERLIN, which was maintained at unity to preserve the high resolution, while adding the needed missing shorter baselines from the VLA to image the faint structure of the radio jet. We re-weighted the data so VLA/$e$-MERLIN contributed the same amount in terms of data weights at the same scales. Using \verb .casa. task \verb .phaseshift.,  we aligned the data into a common phase centre. We then self calibrated the combined data in phase and amplitude.

\begin{figure*}[htbp]
    \centering
    \includegraphics[width=\textwidth]{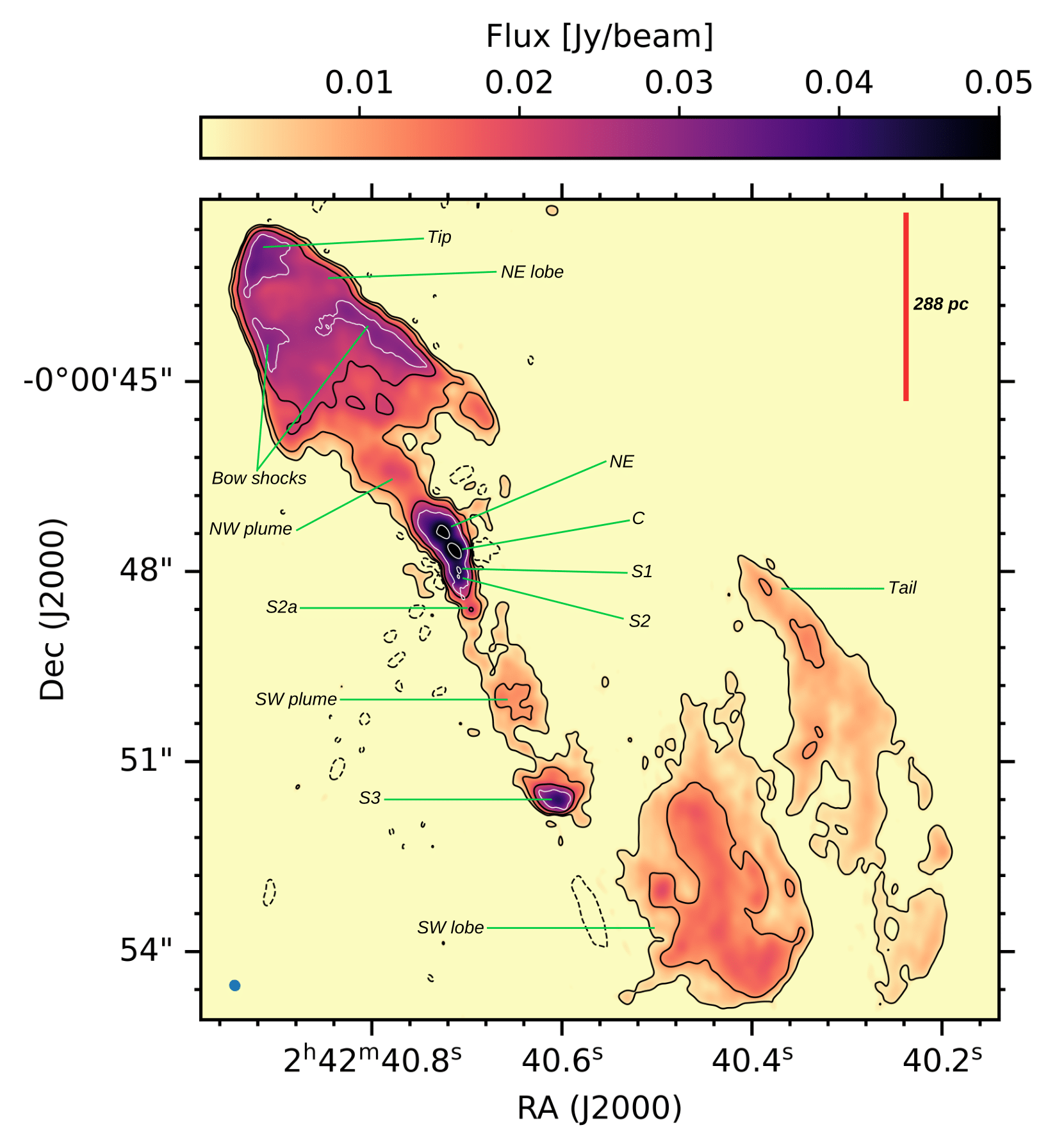}
    \caption{A combined $e$-MERLIN$\Plus$VLA image constructed between 4$-$12 GHz. The map resolves the compact radio components at the central region including the AGN (S1) as well as the extended emission from the NE and SW jet lobes. Black contours show emission at [-3, 3, 9 and 27]$\sigma$ and are imaged with robust 0, yielding an angular resolution of 0.18$\,\times\,0.17$ arcsec. Negative contours are dashed. White contours show emission at 310\,$\sigma$ and are imaged with robust $-$2 yielding angular resolution of 0.12\,$\times$ 0.05 arcsec$^{2}$. The rms sensitivity in both images is $\sigma=30\micro$Jy. The beam for robust 0 data is represented by the bottom left ellipse in blue. The components are labelled according to \citet{Gallimore_1996GTrue} convention.}
    \label{fig:combined-prem}
\end{figure*}

We imaged the data in \verb .casa. with \verb .tclean.. The cell size was set to 0.01 arcsec determined by the $e$-MERLIN data, which has the higher angular resolution of the two datasets. Other imaging parameters were maintained the same as those reported in Section  \ref{vla-4-12-data}. The final images were produced with robustness parameter of 0 and $-$2 to balance between the high angular resolution and extended faint emission. The combined image is shown in Fig. \ref{fig:combined-prem}.

\begin{table*}[htbp]
    \centering
    \resizebox{\columnwidth}{!}{\begin{tabular} {c c c c c c c c c } 
 \hline
 Comp. & Peak flux & Int. flux & Peak flux & Int. flux &Peak flux & Int. flux & Comp. size & Spectral \\
 
 Name & \textit{e}-MERLIN & density & density & density & density & density &  \textit{e}-MERLIN & index \\
      & (mJy\,beam$^{-1}$) & $e$-MERLIN & VLA$-$A$^{\text{X}}$  & VLA$-$A$^{\text{X}}$ & VLA$-$A$^{\text{K}}$ & VLA$-$A$^{\text{K}}$ & (mas$^{2}$) & \\
   &  & (mJy) & (mJy\,beam$^{-1}$) & (mJy) & (mJy\,beam$^{-1}$) & (mJy)&  & \\
 \hline\hline
 NE & $46.6\pm2.5$ & $117.5\pm6.8$   & $16\pm1$ & $57\pm4$    & $10.5\pm0.6$      & $24\pm1$  & $77.5\,\times\,71.4$ & $-1.1\pm0.1$   \\ \hline
 C & $49\pm3$ & $174\pm14$   & $33\pm2$ & $74.4\pm4.8$   & $19\pm1$      & $45.5\pm2.6$   & $84\,\times\,83.6$ & $-0.9\pm0.1$ \\ \hline
 S1 & $9.3\pm0.6$ & $15.2\pm1.2$    & $10.5\pm0.6$ & $24\pm2$    & $12.4\pm0.7$   & $14.7\pm0.8$  & $56.7\,\times\,39.3$ & $0.0\pm0.1$    \\ 
 \hline
 S2 & $9.5\pm0.8$ & $25.6\pm2.6$      & $4.7\pm1.1$ & $13.3\pm4.3$       & $3.1\pm0.2$       & $8.7\pm0.5$ & $62.9\,\times\,53.2$ & $-0.74\pm0.02$     \\ \hline
 S2a & $0.7\pm0.1$ & $0.7\pm0.1$    & $0.56\pm0.03$ & $1.4\pm0.1$   & $0.31\pm0.04$ & $0.7\pm0.1$ & $52\,\times\,15$ & $-0.1\pm0.1$    \\ 
 
 \hline
 S3 & $10\pm1$ & $33.5\pm3.4$     & $3.4\pm0.3$ & $22.4\pm2.1$ & $2.2\pm0.2$      & $9.7\pm1.1$ & $104\,\times\,76$ & $-0.9\pm0.1$ \\ 
 
 \hline
 
\end{tabular}}

\caption{Measurements of the compact radio components of the jet obtained from Fig. \ref{fig:c_k} (a) and (b). The angular resolution and $uv-$coverage of both data were matched to ensure that emission from the same spatial scales were compared in calculating spectral indices as explained in Section \ref{results}.}
\label{table_components}
   
\end{table*}

\section{Results}\label{results}

Our combined image shows all the components identified along the jet with labels (see Fig. \ref{fig:combined-prem}). 
We fitted two-dimensional (2D) Gaussian components to all the compact components (NE, C, S1, S2, S2a and S3) using \verb .casa. task \verb .imfit.. We extracted the integrated fluxes and component sizes as shown in Table \ref{table_components}. We convolved the data with the same circular beam size of 0.06 arcsec$^{2}$ and matched the $uv-$coverage from 50$-$3000 k$\lambda$ to ensure that emission from the same regions of the source were compared in order to calculate a spectral index ($\alpha$). For each component the spectral index is calculated from the integrated flux densities in the 5\,GHz $e$-MERLIN and 21\,GHz VLA images and they are recorded in Table \ref{table_components}. We define spectral index, $\alpha$ as S\,$\alpha$\,$\nu^{\alpha}$, such that negative spectral index corresponds to steep spectra while inverted spectra are positive. Where necessary we measured extended emission from the VLA 6 GHz map in Fig. \ref{fig:vla-C-X} (a), since it is the most sensitive map for the resolved components in the radio jet, and the measured distances are from the combined $e$-MERLIN$\Plus$VLA map in Fig. \ref{fig:combined-prem}.

\subsection{The central region}\label{results:compact}

Component S1, the AGN \citep{Muxlow1996}, is the most compact of all components, both at 5 and 21 GHz. We consider it as the origin of the radio jet \citep[see][]{Muxlow1996, Gallimore_1996GTrue, Gallimore2004_Parsec} and we measure a spectral index of $0.0\pm0.1$ using the $e$-MERLIN and VLA fluxes at 5 and 21 GHz (see Table \ref{table_components}).

Moving north from component S1 is component C which is $\sim$\,0.3 arcsec away. From C the jet diverts by $\sim$\,30$\degree$ to the north-east and $\sim$0.35 arcsecs away,  component NE resides. Component S2 is to the immediate south of the AGN at a distance of $\sim$\,0.12 arcsec. The high angular resolution afforded by the combined image  and a robust parameter of $-$2 was required to separate this component from S1. As shown in \cite{Gallimore_1996GTrue}, a more naturally weighted image could not resolve these components.

We report component S2a for the first time. The component is to the south-west of both the AGN and component S2, at a distance of $\sim$\,0.57 arcsec. It is detected above 3$\sigma$ significance level, with a peak flux of $0.7\pm0.1$\,mJy/beam and $0.31\pm0.04$\,mJy/beam using the $e$-MERLIN data at 5 GHz and the VLA data at 21 GHz, respectively.

\subsection{Monitoring fluxes of central components}\label{monitoring-fluxes}

We compared the peak radio fluxes of all the resolved components at 5 GHz between the 1992 \citep{Muxlow1996,Gallimore_1996GTrue} and our 2018 $e$-MERLIN data. At 21 GHz we compared our 2015 VLA fluxes with the 1983 \citep{Wilson_1983} and 1992 peak fluxes \citep{Gallimore_1996GTrue}. The central frequency of the 1983 and 1992 VLA epochs was 22.4 GHz. We note that the slight difference in frequency does not have a significant impact on the fluxes we observe.

Fluxes for components NE, S1 and S2 were consistent within the error limits for both the 5 and 21 GHz datasets for 1983, 1992 and 2015/2018. However, component C showed a decrease in flux density from 41$\pm$1 mJy to 38.5$\pm$0.7 mJy between 1983 and 1992, and then to 19$\pm$1 mJy in 2015 in the VLA 21 GHz observations. Although the decrease in flux densities between 1983 and 1992 is within uncertainties, the $\sim$\,50\% decrease from 1992 to 2015 is significant, at the 18$\sigma$ level. The $e$-MERLIN 5 GHz peak flux of component C dropped by $\sim$\,20\% from 58$\pm$4 mJy to 49$\pm$3 mJy between 1992 and 2018, a 26-year period corresponding to a significance level of 2$\sigma$.

\subsection{The NE region}\label{result:ne-region}
The NE lobe of NGC\,1068 traces the radio plasma trajectory north-east of component NE. The total extent of the northern component measured between the AGN and jet tip of the NE lobe from Fig. \ref{fig:combined-prem} is $\sim$\,6 arcsec corresponding to 432 pc in projection. The NE lobe is nearly symmetrical with a width of $\sim$\,1.5 arcsec from the central axis measured from the bottom side of the cone-shaped lobe. The western bow shock is 3 arcsec ($\sim$\,240 pc projected) in length and connects to the jet tip. The eastern bow shock in comparison extends $\sim$\,2 arcsec (see Fig. \ref{fig:ne-lobe}).

\subsection{The SW region}\label{result-sw-region}

The SW region connects the components downstream between S2a and the SW lobe. Components S2a and S3 are separated by $\sim$\,3.3 arcsec. The AGN and component S3 are $\sim$\,4.2 arcsec apart. The SW plume resides between S2a and S3. The plume increases in width from $\sim$\,1 arcsec closer to the AGN to $\sim$\,2.5 arcsec as it joins with component S3. The SW lobe is characterised by a region of high surface brightness to the south and a rather low surface brightness \say{tail} extension. The SW lobe spreads to the south and further west of component S3 up to $\sim$\,6 arcsec away. The tail extends about 4 arcsec to the north above component S3 and about 5 arcsec to its west. Its trajectory (see Fig. \ref{fig:combined-prem}) suggests that the lobes may have encountered a higher density medium, resulting in a back flow of the radio plasma.

\section{Discussion}\label{discussion}

\subsection{The central components}\label{diss-central-comp}

\subsubsection{Component NE}\label{diss-comp-NE}

Component NE is the second brightest of the central components of the jet at 5 GHz with both the $e$-MERLIN and VLA (see Table \ref{table_components} and Fig. \ref{fig:c_k}). It is also the furthest compact central component from the AGN in the northern part of the inner jet (see Section \ref{results:compact}). VLBA observations taken in April 1997 at 1.4, 5 and 8.4 GHz show that NE is the brightest compact component in each of the three bands \citep{Gallimore2004_Parsec}. This is likely because the other components are resolved out at VLBA scales and NE has the most extended structure compared with other components, hence collecting more flux in the lower angular resolution VLA and $e$-MERLIN observations presented here. The steep spectral index of $\alpha = -1.1\pm0.1$ is consistent with $\alpha = -$1 and $\alpha = -$1.1, calculated in \cite{Gallimore_1996GTrue} and \cite{Gallimore2004_Parsec} respectively. Such a steep spectral index can indicate that the emission is from optically-thin non-thermal emission.

\cite{Gallimore2004_Parsec} suggests that the blob at component NE could be a result of local density enhancement in the jet flow originating from the jet base due to variable accretion rates at the AGN, which may in turn result in variable outflow speeds, leading to internal shocks (e.g \citealt{blandford1979}). While this view point is likely, recent ALMA observations of CO (3$-$2) and C$_2$H (1$-$0) transitions in \cite{Garcia-Burillo2014} and \cite{Garcial_alma_imaging_2017A&} respectively shows that the entire central region of the jet lies within a dense molecular ISM region that occupies the inner $\sim$\,4 arcsec of the galaxy (e.g \cite{viti_line_ratio_2014A, Garcia-Burillo2014,Garcial_alma_imaging_2017A&}), referred to as the circum-nuclear disc (CND), a region of complex physical and chemical activities, rich in dense molecular gas. We therefore can not rule out the possibility of jet$-$cloud interaction at NE.

\subsubsection{Component C} \label{diss-comp-C}

Component C was originally thought to be the site of the accreting SMBH, as it is the brightest component at VLA and $e$-MERLIN observations, until this was disproved by \cite{Muxlow1996}, who showed that the AGN resides in component S1 (see next section). The spectral index of component C is $-0.9\pm0.1$. The steep spectral index that we measured for this component is an indicator of synchrotron emission within the jet. Our findings indicate a difference in flux between the 1992 and 2015 VLA epochs at 21 GHz with 18$\sigma$ significance level, a decrease by 50\%. A similar trend is also observed with the 1992 and 2018 $e$-MERLIN epochs at 5 GHz where the flux drops by 20\%, with 1.8$\sigma$ significance level (Section \ref{monitoring-fluxes}). We note that the flux has varied significantly, which could explain the difference in spectral index between the two epochs.

Fluxes from our 2015 and 2018 $e$-MERLIN and VLA data for component C have decreased. We took into account other factors that can cause these differences, such as different or variable flux calibrators or different weighting schemes used in image deconvolution processes. We found that the same flux calibrator, 3C348 was used in all epochs and that it was not varying over time. The weighting schemes used did not amount to any significant differences in fluxes in other components. Furthermore, other components show consistency in fluxes across the epochs. We took this as evidence that the flux variations in component C are not as a result of any measurement, deconvolution or calibration errors. It is possible that component C is varying due to continuous powerful injections of relativistic particles within a dense molecular cloud, causing evolution of the strong jet shock.

The VLBA observations of this component by \cite{Parsec_2_Gallimore_2004} showed that the flux of component C had gone down from 27\,mJy in May 1996 to 19\,mJy in April 1997, suggesting variability on less than a year timescales (331 days). The variability timescale implied a resolved source of angular diameter of $\leq$ 0.008 arcsec in size,  corresponding to $\leq$ 0.6 pc. The source variability could be caused by evolving jet$-$ISM shocks caused interaction between the jet and molecular clouds, evidenced from H$_{2}$O observations \citep{maser_Greenhill_1997,Gallimore_nuclear_watermaser_2001,Parsec_2_Gallimore_2004}.

Our new $e$-MERLIN observations (project ID: CY13006) in six epochs spread over eight weeks have been obtained to investigate the short term variability of this compact component as well as the supernova SN2018ivc. The results of these data will be analysed and featured in future works.

At component C, the jet diverts by $\sim\,30\degree$ to the north-east (see Fig. \ref{fig:combined-prem}). The cause for deviation is a dense gas cloud that is on the jet's path \citep{Gallimore2004_Parsec}. Further evidence of dense molecular clouds at component C are the associated H$_{2}$O masers \citep{maser_1996_gallimore_1996, Gallimore_nuclear_watermaser_2001, maser_Greenhill_1997,masers-alma-morishima,gallimore-masers-2023}. ALMA HCO$^+$ (3$-$2) and HCN (3$-$2) transitions \citep{imanishi-2016-gas}, suggest the presence of dense gas at component C. However, the 3D geometry nature of the galaxy makes it hard to confirm if this dense cloud is within the same region as component C.

\subsubsection{Component S1}
Component S1 was first suggested to be the AGN by \cite{Muxlow1996}. The flat radio spectrum is associated with non-thermal emission from continuous particle injection, expected from the AGN (see Table \ref{table_components}). In recent works by \cite{gallimore-masers-2023}, observations of blue and red shifted nuclear H$_{2}$O masers rotating about component S1 with a calculated central mass of 1.7$\times10^{7}$ $M_{\odot}$ with the Very Long Baseline Array (VLBA), further shows that S1 is the AGN. In this work, we report a spectral index of $0.0\pm0.1$ for S1. 

Hard X-ray observations (2$-$10 keV) show that the brightest point is coincident with S1 \citep{chadra_xraY_young_2002}. An Fe K$\alpha$ line was detected at 6.4\,keV, indicative of reflection of X-ray photons by the torus around the AGN \citep{mouchet-Fe-agn-2001}, further suggesting S1 as the AGN.

\subsubsection{Component S2}

Component S2 lies south of the AGN and forms the start of the southern jet component. It is the closest compact component to the AGN at a distance of $\sim$\,0.12 arcsec ($\sim$\,8.6 pc). We calculate $\alpha = -0.74\pm0.02$ for this component. This is indicative of synchrotron radiation from a relatively old relativistic plasma. The spectrum is flatter at VLBA scales, $\alpha = -$0.15$\pm0.08$ between 5 and 21 GHz \citep{Parsec_2_Gallimore_2004}. The VLBA data is resolves more flux than $e$-MERLIN, giving component sizes of $\sim$\,0.006 arcsec and 0.06 arcsec respectively. This explains why the spectral index measured between the $e$-MERLIN and VLBA scales is different. The VLBA spectral index is flatter as there is a localised region where more relativistic particles are hitting the gas clouds. The larger beam size of $e$-MERLIN collects additional old synchrotron emission from the extended region of the component, steepening the spectral index compared to VLBA fluxes.

\subsubsection{Component S2a}

We label this unresolved component S2a in Fig. \ref{fig:combined-prem}, which has not been reported before. The component is detected with peak flux of $0.7\pm0.1$\,mJy/beam. With an almost flat spectrum of $\alpha\,=\,-0.1\pm\,0.1$. Component S2a is located in a region with dense molecular gas within the CND \citep{viti_line_ratio_2014A}. It is likely that this component is created as a result of jet$-$cloud impact along the southern jet.

\subsection{The extended jet structure}\label{diss-extend-structure}

\subsubsection{The NE lobe}

\begin{figure}[!ht]
\centering
    \subfloat[\centering]{{\includegraphics[width=8.5cm]{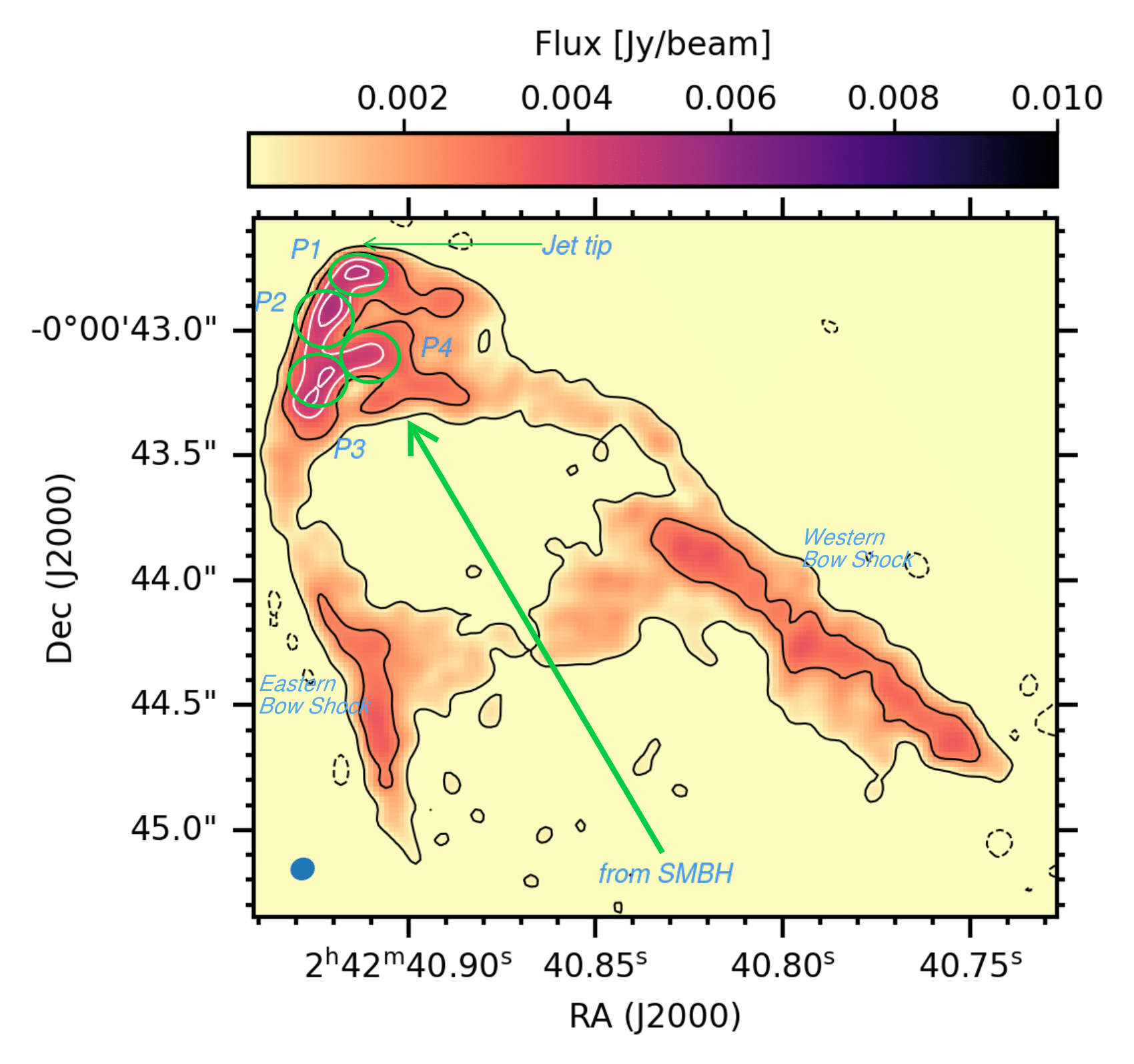}}}%
    \qquad
    \subfloat[\centering]{{\includegraphics[width=8.5cm]{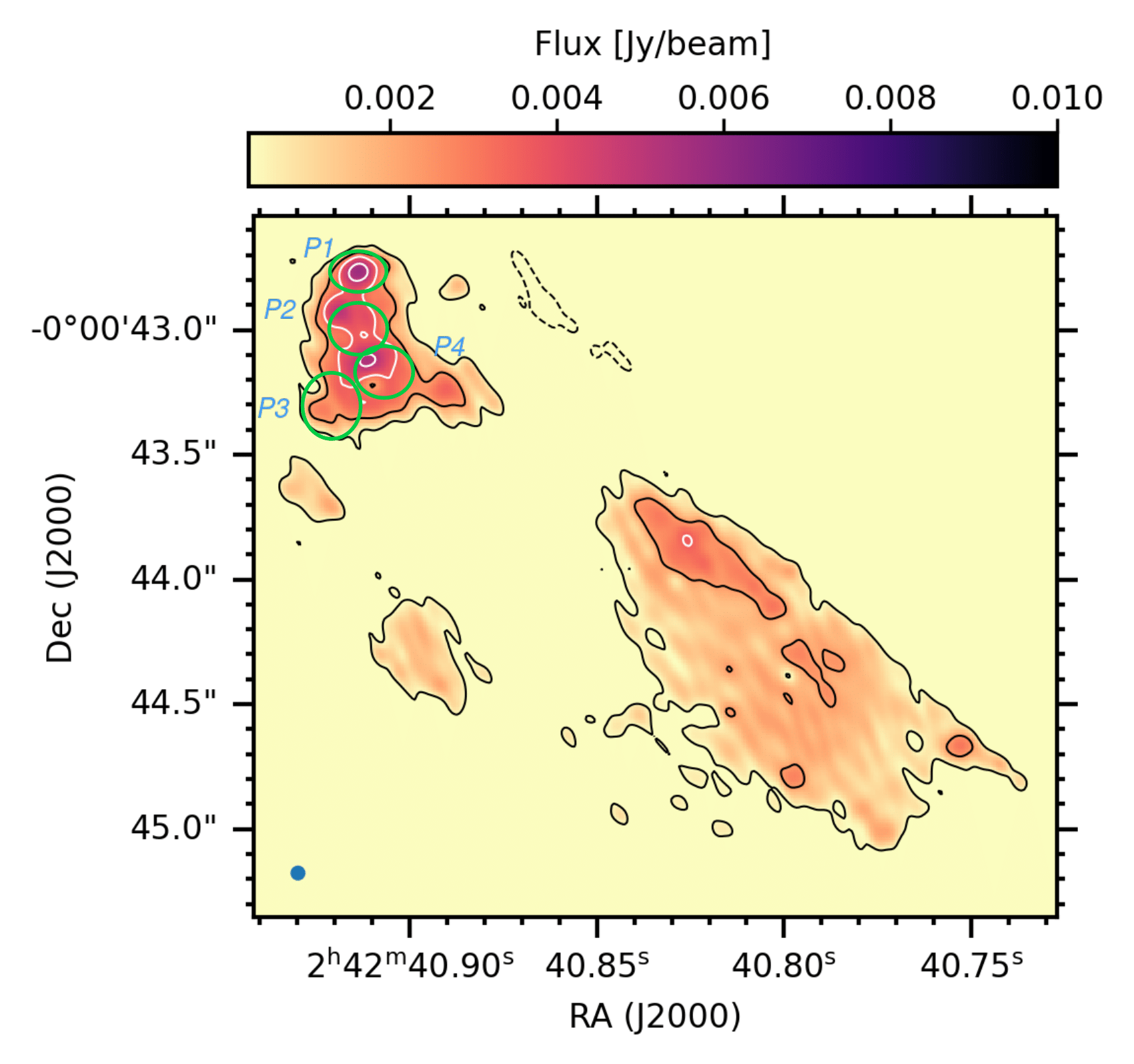}}}%
        \caption{(a) The VLA 21 GHz image of the NE lobe, created with Briggs weighting, robust 0. Contour levels are; black [-7, 10 and 30]$\sigma$, white [60 and 90]$\sigma$. The rms noise levels are 1 $\micro$Jy and are represented by $\sigma$. The clean beam size is 0.10 $\times$ 0.09 arcsec$^{2}$ and is represented by the blue patch in the bottom left corner. (b)$e$-MERLIN 5 GHz image (zoomed in from Fig. \ref{fig:c_k} (a)) of the jet lobe head containing the hotspots P1$-$P4. Contour levels are; black [-7, 5 and 10]$\sigma$ and white [16 and 28]$\sigma$. $\sigma = 6\micro$Jy. The beam size is 0.06 arcsec, represented by the blue path in the bottom left corner. In both images, negative contours are dashed.}
    \label{fig:ne-lobe}
\end{figure}

The jet cuts through a region of dense molecular gas north of the central region, creating the NE limb-brightened lobe with bow shocks that define the edges of an outflow \citep[see Figs. \ref{fig:c_k} and \ref{fig:ne-lobe} and][]{Garcia-Burillo2014,Garcial_alma_imaging_2017A&}. The bow shocks are formed from the transfer of kinetic and radiative energy into the ISM \citep{wilson_bowshocks,Bowshocks_Taylor_NLR_1992,Jets_bowshock_Tram_2018}. The shocked regions propagate sidewards relative to the jet, producing radio emission (see Fig. \ref{fig:ne-lobe}). Dense molecular gas outflows are detected in a cone-shaped region coinciding in position (in 2D geometry) with the NE lobe \citep{Garcia-Burillo2014,Garcial_alma_imaging_2017A&}. They suggest that the NE jet lobe is driving the molecular gas out of this region.

\cite{Garcia-Burillo2014} calculated molecular outflow rates of $\sim\,63\,M_{\odot}$\,yr$^{-1}$ within the entire CND region (area of radius of $\sim 4 \arcsec$ around the AGN, including the NE lobe region) and $\sim\,6\,M_{\odot}\,$yr$^{-1}$  around the bow-shock arc region (coincident with the NE lobe). Therefore, the outflow rate of the NE lobe is $\sim$9.5$\%$ of the entire CND region. Using this ratio, we consider that about 9.5$\%$ of the kinetic power, $L_{kin}\,\sim\,5\,\times\,10^{41}$\,erg\,s$^{-1}$, calculated from the entire CND \citep{Garcia-Burillo2014}, is from the NE lobe. This gives a value of $L_{kin-NElobe}\,\sim\,4.8\,\times\,10^{40}$\,erg\,s$^{-1}$.

The monochromatic luminosity ($L_{radio}$) of the lobes at 5 GHz in the VLA image (see Fig. \ref{fig:vla-C-X} (a)) can be computed using the relation \citep{Birzan_2008},

\begin{equation}
    L_{radio} = 4 \pi\,\times\,d^{2}\,\times\,\textit{(}1\Plus z\textit{)}^{\alpha - 1} \times S_{\nu_{0}} \times \nu_{0},
\end{equation}

where d is distance to the galaxy, $z$ is the redshift, $\alpha$ is the spectral index, $S_{\nu_{0}}$ is the flux density and $\nu_{0}$ is the observing frequency. A value of  $\alpha$\,$=$\,$-$0.6 was estimated from the total integrated fluxes of the NE lobe between the VLA 6 and 20 GHz data. To do this, we drew a box around the cone-shaped region in our VLA maps, and extracted the total fluxes with CASA task $imfit$. This value is also consistent with other literature, e.g \citep{Wilson_1983,michi-NW-lobe-2022}. Our calculations of $L_{radio}$ gave a value of $\,$4.5\,$\times\,$\,10$^{38}$\,erg\,s$^{-1}$. We related the radio jet power, $P_{jet-NElobe}$ with $L_{radio}$ as follows in \citealt{cava_feedback_outflow_2010},

\begin{equation}\label{eqn2}
  P_{jet} = 5.8\,\times\,10^{43} \left(\frac{L_{radio}}{10^{40}}\right)^{0.7} erg\, s^{-1}, 
\end{equation}

Therefore, we calculate the NE lobe jet power, $P_{jet-NElobe}\,=\,$6.7\,$\times\,$\,10$^{42}$\,erg\,s$^{-1}$. Using the same method to estimate the total jet power of the entire jet structure, \cite{Garcia-Burillo2014} estimated $P_{jet}\,=\,$1.8\,$\times\,$\,10$^{43}$\,erg\,s$^{-1}$ using the compact radio components NE, C, S1 and S2 totalling a flux density of 840 mJy from the 1.5 GHz MERLIN map from \cite{Gallimore_1996GTrue}. Our calculations focus on the NE lobe, which has a total flux density of 367 mJy, enabling us to relate the kinetic power of the outflows to the radio luminosity of the NE lobe.

According to \cite{Garcia-Burillo2014}, the jet power must be at least 30 times higher than the kinetic power in order to cause molecular outflows, i.e $\frac{P_{jet}}{L_{kin}}>30$. Relating our $P_{jet-NElobe}\,=\,$6.7\,$\times\,$\,10$^{42}$\,erg\,s$^{-1}$ and $L_{kin-NElobe}\,\sim\,4.8\,\times\,10^{40}$\,erg\,s$^{-1}$ we find that the NE lobe has sufficient radio power to cause molecular outflows within the bow-shock arc region even when low coupling efficiency is assumed \citep{Garcia-Burillo2014}.

Recent ALMA observations of the radio jet lobe, at 92 GHz, show similar structure as our data \citep{michi-NW-lobe-2022}. The hotspot components P1$-$P4 with spectral indices of $-0.50\pm$0.16, $-0.59\pm$0.14, $-0.65\pm$0.10 and $-0.50\pm$0.09 respectively between 15 GHz (VLA) and 250 GHz (ALMA)  (discussed in \citealt{michi-NW-lobe-2022}) are detected in our VLA 21 GHz image (see Fig. \ref{fig:ne-lobe} (a)) and $e$-MERLIN 5 GHz image (Fig. \ref{fig:ne-lobe} (b)). Hotspots can be created by shock-driven interactions at the jet-ISM boundary, as evidenced in our radio observations (see simulations by \citealt{hots-spots-hardcastle}).

\subsubsection{Component S3}

Component S3 is the brightest known component in the southern jet. It is detected in all our high angular resolution radio observations of this jet. With a steep spectral index, $\alpha = -0.9\pm0.1$, it implies that non-thermal emission dominates the component, and it is likely due to old jet emission. After component S3, the southern jet diverts at an angle of $\sim$\,30$\degree$ to the west. ALMA maps by \cite{Garcia-Burillo2014} show a dense gas cloud potentially coinciding in location with component S3. We presume that the compact component identified in the ALMA maps is responsible for the radio emission at component S3, and is the cause of the diversion of the jet. We will explore this possibility in a future work.

\subsubsection{The SW lobe}

The complex orientation of the southern jet, behind the galactic plane makes it harder to determine the causes of the emission. A number of possibilities can explain the radio emission from the SW lobe.

First, it is possible that star formation can partially explain the radio emission. Radio emission in 1.5 GHz VLA images \citep{Gallimore_1996GTrue} shows that a star-burst (SB) bar connects from the spiral arms to the CND. The radio jet of NGC 1068 is oriented along the SB bar \citep{Garcia-Burillo2014, Gallimore_1996GTrue}. Therefore, star formation in the SB bar could possibly explain some of the radio emission.

Second, the emission could be relic, coming from the radio jet in the past activity and marking different era's of the AGN activity. This can only be tested by spectral ageing analysis of the jet, a task we aim to carry out in the next paper.

However, it is most plausible that the emission is produced by a jet deflected by dense molecular clouds at component S3 (see \cite{Garcia-Burillo2014}, where the location of S3 is coincident with dense molecular clouds). The deflected southern jet then travels west of component S3. The jet connects directly to the rest of the SW lobe structure through component S3 at higher contour levels, as shown in Fig. \ref{fig:vla-C-X}. This is also demonstrated with VLA A\,-\,array 1.5 GHz (L band) images of NGC 1068 by \cite{Gallimore_1996GTrue} and VLA B\,-\,array 5 GHz images by \cite{ho-seyferts-2001}. The presence of component S3 would cause jet diversion as long as it is dense and massive enough to overcome the jet's kinetic power \citep{Gallimore2004_Parsec}. The impact would cause the jet to spread out as it travels through the starburst region. ALMA 92\,GHz images of the jet by \cite{michi-NW-lobe-2022} shows the SW lobe connecting to component S3.

\subsection{Comparison with other nearby Seyfert galaxies}

About 10\% of all the observable galaxies in the Universe are Seyfert galaxies \citep{seyferts-survey-1995ApJ}. These sources are part of a larger class of galaxies that host an AGN. Our images show that the SMBH in NGC 1068 is actively accreting material and producing the radio jets (about 0.5 kpc from the SMBH) shown in Fig. \ref{fig:combined-prem}. In more powerful radio sources such as 3C 465 ($z$ = 0.03035) the radio jets extend well beyond 150 kpc from the AGN \citep{emmanuel-jets-2020}; about 30 times larger than the radio jets in NGC 1068. Seyfert galaxies are low power AGN, some with radio jets that lie within their galactic bulge, shaping the dynamics of ISM within the galaxy, as is the case with NGC 1068 and Circinus galaxy \citep{Israel-circinu-ism-1992,elmou-circinus-1996,agn_radio_emission_panessa_2019}.

However, not all Seyfert galaxies have well-defined radio jets. For example, a VLA survey of galaxies by \cite{ho-seyferts-2001}, showed that while NGC 1068, NGC 6951, NGC 358, NGC 3031 and NGC 3516 had well-defined jet-like extensions and lobes, NGC 3079 had an irregular structure within the core that could not be easily defined as a jet. Another survey by \cite{Lemmings-II-baldi} found that out of 13 Seyfert galaxies detected in the radio, 6 had a core/core-jet and only 1/13 showed large scale jets akin to those in NGC 1068. About 44\% (19/43) of radio quiet AGN in a survey by \cite{gallimore-2006AJ} were found to be Seyfert galaxies that hosted a kilo-parsec scale radio (KSR) outflows. Such structures can be detected when interferometric observations are not artificially suppressed by leaving out short baselines (VLA B, C and D arrays) hence biasing towards nuclear dominated jet structures. Such KSRs may be connected to nuclear jet emission as is the case with NGC 1068 where the extended jet lobes on kpc scales join the nuclear jets when observed with higher resolution interferometers (see Fig. \ref{fig:c_k}). For Seyfert galaxies that have radio jets, some have jet lobes such as NGC 1068 (the NE and SW such as shown in Fig. \ref{fig:combined-prem}), while others like NGC 4151 show individual components in a line \citep{willliams-ngc4151}. Other radio quiet AGN like NGC 1377 have been discovered to have apparent radio jets with collimated jet-like molecular outflows \citep{saalto-ngc1377}.

Seyfert galaxies have been reported to show radio flux variability in their cores in the past. The radio flux variation in galactic cores is in some cases caused by flares of ionised material being ejected from the accretion disc. In an 8 GHz VLA survey of eleven Seyfert galaxies by \cite{mundell-varying-seyferts-2009}, five Seyferts (NGC 2110, NGC 3081, MCG −6-30-15, and NGC 5273) showed decreases in their nuclear flux between 1992 and 1999, while one (NGC 5273) showed an increase. Core fluxes of six Seyfert galaxies (Mrk 607, NGC 1386, Mrk 620, NGC 3516, NGC 4968, and NGC 7465) remained unchanged over the same time duration. VLBI observations of NGC 2992 at 5 GHz reported core radio flux variation, recording a decrease by a factor of $>$ 3 with within a region of $<$ 0.02 pc \citep{frame-X-II_2022}. We did not find any noticeable change in AGN flux in NGC 1068 in all the epochs we have. However, we cannot rule out AGN variability on short time-scales as our observations are more than 23 years apart.

Radio jets can interact with nearby ISM, creating radio blobs whose flux may vary with time, as more relativistic particles from the SMBH are injected into the shocked regions. For example, component C4E in NGC 4151 \citep{mundell-4151-2003,willliams-ngc4151} has been reported to increase in flux over a period of 22 years from $37.14\pm0.07$ mJy to $66.76\pm0.04$ mJy \citep{willliams-ngc4151}. Here, we have reported the decrease in flux of component C in NGC 1068 from 38.5$\pm0.7$ mJy in 1992 to 19\,$\pm1$\,mJy in 2015 with significance level of 18$\sigma$, over a 23-year period. It is difficult to understand the variation patterns within such long time periods as we have. More observations may need to be conducted over shorter time durations to better understand component C.

\section{Summary and Conclusions}

We have used high sensitivity $e$-MERLIN 5 GHz and \mbox{VLA 4 $-$ 23.5\,GHz} observations to map the radio emission from the radio jet in NGC 1068. The jet lies in the inner 4 arcsec of the galaxy, referred to as the circum-nuclear disc (CND). For the first time, we created a combined 5 GHz $e$-MERLIN and VLA map of NGC 1068, tracing the compact central components and the extended structure of the jet lobes.

We summarise and conclude our key findings as follows:
\begin{enumerate}[wide, labelwidth=!,itemindent=0pt,labelindent=0pt, leftmargin=1.5em, label=\roman*,itemsep=0.2cm]
    \item The compact central components NE, C, S1 and S2 to the extended structure of the NE and SW jet lobes have been shown in our combined $e$-MERLIN and VLA data at spacial scales of 15\,k$\lambda$\,$-$\,3300\,k$\lambda$.

    \item We report the discovery of component S2a detected above 10$\sigma$ with both the $e$-MERLIN and VLA.
    
    \item Components NE, C and S2 have a steep spectrum of $-1.1\pm0.1$, $-0.9\pm0.1$ and $-0.74\pm0.02$ respectively, indicative of optically-thin non-thermal emission between 5 and 21 GHz. Component S1 has a nearly flat spectrum of, $0.0\pm0.1$ and is the AGN.
    
    \item Component C has significantly decreased in flux by 18$\sigma$ between 1992 and 2015: At 21 GHz the VLA flux decreases from 38.5$\pm$0.7\,mJy to 19$\pm$1\,mJy in this time period. At 5 GHz with $e$-MERLIN, the flux decreases from 59$\pm$4\,mJy to 49$\pm$3\,mJy between 1995 and 2018. High angular resolution VLBA results of the resolved component C by \cite{Parsec_2_Gallimore_2004} reported flux decrease from 27\,mJy to 19\,mJy in 1996 within a period of 331 days. Our findings agree with conclusions made by \cite{Parsec_2_Gallimore_2004}, that component C is varying.
    
    \item We detect the bow shocks in the north-east jet lobes that coincide with the molecular gas outflows observed with ALMA. The north-east jet lobe has a jet power of $P_{jet}\,=\,$6.7\,$\times$\,10$^{42}$\,erg\,s$^{-1}$ believed to be responsible for driving out dense molecular gas of mass, $\sim\,9\,\times\,10^{6}$\,$M_{\odot}$ \citep{Garcia-Burillo2014} around itself.
\end{enumerate}

 This work demonstrates the capabilities of $e$-MERLIN and VLA to image the complex radio structure of NGC\,1068 across a wide range of spatial scales. It remains unclear whether the emission in the SW lobe is due to the jet or from star formation activity in the same region. In future studies, we aim to disentangle the nature of the emission along the jet by calculating the spectral age of the electrons and by comparing to multi-wavelength data of similar resolution, i.e. ALMA, to align the radio emitting regions and molecular gas in this galaxy.
 
\section*{Acknowledgements}

We thank the anonymous reviewer for their comments and revisions. We thank the Technical University of Kenya (TUK) and Development in Africa with Radio Astronomy (DARA) for funding this research. The Jodrell Bank Center for Astrophysics (JBCA), University of Manchester, for expert and computational support offered though out this work. We thank the $e$-MERLIN and VLA for the observations. $e$-MERLIN is a National Facility operated by the University of Manchester at Jodrell Bank Observatory on behalf of STFC. The VLA is operated by the NRAO.

\section*{Data Availability}

The data on which this paper is based are publicly available from the e-MERLIN and VLA archive under projects IDs described in Section \ref{observations}. Calibrated image products are available upon reasonable request to the corresponding author.

\bibliographystyle{mnras}
\bibliography{example} 

\bsp
\label{lastpage}
\end{document}